\documentclass[twocolumn,superscriptaddress,prb,10pt,nofootinbib]{revtex4-1}
\usepackage{braket}
\usepackage{amsmath,amssymb}
\usepackage{tikz}
\usepackage{graphicx}
\usepackage{color}
\usepackage[colorlinks,bookmarks=false,citecolor=blue,linkcolor=red,urlcolor=blue]{hyperref}
\usepackage{times}

\pdfoutput=1

\begin{document}

\title{Quench action and R\'enyi entropies in integrable  systems} 

\author{Vincenzo Alba}
\author{Pasquale Calabrese}
\affiliation{International School for Advanced Studies (SISSA),
Via Bonomea 265, 34136, Trieste, Italy, 
INFN, Sezione di Trieste}

\date{\today}

\begin{abstract} 

Entropy is a fundamental  concept in equilibrium statistical mechanics, yet its origin
in the non-equilibrium dynamics of isolated quantum systems is not fully understood. 
A strong consensus is emerging around the idea that the stationary thermodynamic entropy is 
the von Neumann entanglement entropy of a large subsystem embedded in an infinite system. 
Also motivated  by cold-atom experiments, here we consider the generalisation to R\'enyi entropies.  
We develop a new technique to calculate the {\it diagonal} R\'enyi entropy in the quench action formalism. 
In the spirit of the replica treatment for the entanglement entropy, the diagonal R\'enyi entropies are generalised 
free energies evaluated over a thermodynamic macrostate which  depends on the R\'enyi index and, in particular,
it is not the same describing the von Neumann entropy.
The technical reason for this, maybe surprising, result is that the evaluation of the moments of the diagonal density matrix 
shifts the saddle point of the quench action.  
An interesting consequence is that different R\'enyi entropies encode information about different regions of the spectrum of 
the post-quench Hamiltonian. 
Our approach provides a very simple proof of the long-standing issue that, for integrable systems, the diagonal 
entropy is half of the thermodynamic one and it allows us to generalise this result to the case of arbitrary R\'enyi entropy.

\end{abstract}

% \pacs{73.43.Cd, 71.10.Pm  {\tt check!}}

\maketitle

%#############-- INTRODUCTION --########################################

{\bf Introduction.}--An intriguing question for an out-of-equilibrium isolated system evolving from a pure state  $|\Psi_0\rangle$ is 
to understand microscopically the thermodynamic entropy of the emergent statistical ensemble. 
Indeed, it has been undoubtedly established that for large times the local properties of  a quenched system are correctly captured by 
a statistical ensemble which is either the Gibbs or generalised Gibbs depending on whether the system is generic ~\cite{v-29,eth0,deutsch-1991,srednicki-1994,rdo-08,rigol-2012,dalessio-2015} 
or integrable~\cite{rigol-2007,cazalilla-2006,barthel-2008,cramer-2008,fagotti-2008,cazalilla-2012a,
calabrese-2012,sotiriadis-2012,collura-2013,fagotti-2013,p-13,fe-13b,
kcc14,sotiriadis-2014,fcec-14,ilievski-2015a,alba-2015,langen-15,essler-2015,cardy-2015,
sotiriadis-2016,bastianello-2016,vernier-2016,pvw-17,vidmar-2016,gogolin-2015,
calabrese-2016,ef-16}  respectively. 
However, since the entire system is always in a zero-entropy pure state because of the unitary time evolution,
it is not obvious whether the entropy of the statistical ensemble has any physical meaning. 
This apparent paradox has been nowadays clarified and it is widely accepted that the thermodynamic entropy is \cite{foot1} the stationary value 
of the entanglement of a large subsystem $A$ embedded in an infinite system \cite{calabrese-2005,spr-12,dls-13,collura-2014,kormos-2014,bam-15,kaufman-2016,cc-16,alba-2016,d-17}. 
This entanglement is quantified by the von Neumann entropy $S_A\equiv-\textrm{Tr}\rho_A\ln\rho_A$ of the reduced density 
matrix $\rho_A$ of the subsystem $A$.
Very recently, the equivalence between stationary entanglement entropy and thermodynamic entropy has been 
exploited to provide analytic exact predictions for the entire time dependence of the entanglement entropy in integrable models \cite{alba-2016}
as carefully tested against numerical simulations in the XXZ spin-chain \cite{alba-2016}.
% and generalised to other integrable models \cite{mbpc-17,ac-prep}.

An alternative and natural definition of entropy after a quench is the diagonal entropy $S_d$~\cite{polkovnikov-2011}, 
which is the von Neumann entropy $S_d=-\textrm{Tr}\rho_d\ln\rho_d$ of the diagonal ensemble 
\begin{equation}
\label{rho-diag}
\rho_{d}=\sum\limits_{m=1}^\infty w_m|m\rangle\langle m|\quad\textrm{with}\quad
w_m\equiv |\langle\Psi_0|m\rangle|^2.
\end{equation}
Here the states $|m\rangle$ are the eigenstates of the post-quench Hamiltonian $H$ and the coefficients 
$w_m$ are the overlaps with the initial state $|\Psi_0\rangle$. 
The great advantage of the diagonal entropy is that  can be computed  
from the  initial state, without solving the complicated quench dynamics.  
A lot of effort has been devoted to understand the relation between the diagonal entropy and
the stationary subsystem entanglement entropy (which, as already stressed, is the entropy of the stationary ensemble). 
It has been suggested that for generic systems relaxing to a Gibbs ensemble the diagonal entropy is the same as the thermodynamic 
one \cite{polkovnikov-2011}, as tested also in numerical simulations \cite{santos-2011,spr-12}.
Contrarily, for quenches to integrable models it has been found numerically and analytically 
that the diagonal entropy is half of the thermodynamic one \cite{spr-12,gurarie-2013,fagotti-2013b,dora-2014,collura-2014,kormos-2014,piroli-2016}.
A solid physical argument to explain this factor $1/2$ has been provided by Gurarie  \cite{gurarie-2013}, but so 
far there was no proof, neither a complete understanding of its generality.

For a generic density matrix $\rho$, the R\'enyi entropies
\begin{equation}
\label{d-renyi}
S^{(\alpha)}\equiv\frac{1}{1-\alpha}\ln
\textrm{Tr}\rho ^\alpha\,, % =\frac{1}{1-\alpha}\ln\sum\limits_m w_m^\alpha, 
\end{equation}
represent a more general class of entropies generalising the von Neumann one.
Here $\alpha$ is the index of the R\'enyi entropy and it is an arbitrary positive real number. 
In the limit $\alpha\to1$, \eqref{d-renyi} gives back the von Neumann entropy. 
Although R\'enyi entropies for generic $\alpha$ do not have some important properties 
of the von Neumann one (the most physical one being the strong subadditivity),  they are 
attracting a lot of attention, especially in connection with entanglement.
Indeed, R\'enyi  entanglement entropies (i.e. \eqref{d-renyi} with $\rho=\rho_A$) are 
entanglement monotones \cite{v-00}, i.e.  quantities 
that do not increase under local operations and classical communication which is the 
technical requirement for any good measure of entanglement. 
Furthermore,  for integer $\alpha$, they are the core of the replica approach to the entanglement entropy \cite{cc-04}. 
While the replica method was introduced mainly as a theoretical trick \cite{cc-04}, it became a useful tool  
for the experimental measurement of entanglement: R\'enyi entanglement entropies (for $\alpha=2$) have been measured 
experimentally with cold atoms, both in equilibrium~\cite{islam-2015} and after a quantum quench~\cite{kaufman-2016}.

The main goal of this paper is to initiate the investigation of R\'enyi entropies in interacting integrable models  
also, but not only, to provide predictions testable in experiments.
To this aim, we adapt the Thermodynamic Bethe Ansatz (TBA) techniques for quantum quenches (overlap TBA or Quench Action 
method~\cite{caux-2013,caux-16}) and we derive analytically the diagonal R\'enyi entropies (i.e.~\eqref{d-renyi} with $\rho=\rho_d$). 
In order to make this paper as readable as possible, we report only the general framework for calculating
the diagonal entropies and we discuss in details the main physical consequences of our findings. 
We postpone a detailed analytical and numerical technical analysis of the TBA solutions for specific integrable models to 
forthcoming publications \cite{ac-prep}.

{\bf The Thermodynamic Bethe ansatz (TBA)}.--In Bethe ansatz  solvable models, the eigenstates are in correspondence with a 
set of pseudomomenta $\lambda$, usually called rapidities. 
These rapidities satisfy a set of non-linear coupled equation known as Bethe equations. 
In the thermodynamic limit \cite{foot2} the rapidities form a continuum and the eigenstates are 
characterised by the densities of particles $\pmb{\rho}(\lambda)$ and of holes $\pmb{\rho}_h(\lambda)$ (holes are unoccupied Bethe modes that 
in an interacting model are not simply related to $\pmb{\rho}$).
With bold symbols $\pmb{\rho}$ and $\pmb{\rho}^{(h)}$ we 
denoted the full set of all stable Bethe quasi-particles, i.e. $\pmb{\rho}\equiv\{\rho_n\}_{n=1}^{\cal N}$,
where ${\cal N}$ depends on the considered model  (e.g. there is only one species 
for the repulsive Lieb-Liniger model, but  infinite for the gapped XXZ chain). 
In a given eigenstate, the densities $\pmb{\rho}$ and $\pmb{\rho}^{(h)}$ are %not independent, but they are
related by the continuum limit of the Bethe equation.
A set of densities identifies a thermodynamic macro-state that corresponds to an 
exponentially large number of microscopic eigenstates. 
The total number of microstates with the same $\pmb{\rho}$ and $\pmb{\rho}^{(h)}$ is $e^{S_{YY}}$, with $S_{YY}$ 
the thermodynamic Yang-Yang  entropy  of the macrostate \cite{yy-69}
\begin{multline}
\label{y-y}
S_{YY}[\pmb{\rho}]\equiv L\sum_{n}\int d\lambda\Big[\rho_n^{{}_{(t)}}\ln\rho_n^{{}_{(t)}}\\ 
-\rho_n\ln\rho_n-\rho_n^{{}_{(h)}}\ln\rho_n^{{}_{(h)}}\Big],
\end{multline}
where  the total density is $\rho_n^{{}_{(t)}}\equiv \rho_n+\rho_h^{(h)}$.
For systems in thermal tequilibrium, $S_{YY}$ is the thermal entropy~\cite{taka-book}. 

In the thermodynamic limit, a sum over eigenstates 
is replaced by a functional integral over the rapidity densities as
\begin{equation}
\label{rep}
\sum_m\to\int{\mathcal D}\pmb{\rho}e^{S_{YY}[\pmb{\rho}]}, 
\end{equation}
where the factor $e^{S_{YY}}$ counts the exponentially large number of microscopic eigenstates leading to the 
same densities.

{\bf The quench action}.--The out of equilibrium dynamics of integrable systems can be cast in the TBA language with  the quench action
method~\cite{caux-2013,caux-16}.
Although this approach can be  used  to study the time evolution (as done in a few cases \cite{dc-14,dlc-15,pc-17}), 
it mainly provides a calculable and manageable representation of the stationary state (or equivalently of the diagonal ensemble),
as nowadays explicitly worked out for many integrable models \cite{BeSE14,BePC16,pozsgay-13,wdbf-14,PMWK14,ac-16qa,ppv-17,mbpc-17,dwbc-14,pce-16,Bucc16,npg-17}.
The stationary value of a local observable ${\cal O}$, i.e. its value in the diagonal ensemble, is
\begin{equation}
\label{d-ensemble}
\langle{\mathcal O}\rangle_d=\sum_m|\langle\Psi_0|m
\rangle|^2\langle m|{\mathcal O}|m\rangle.
\end{equation}
According to the recipe \eqref{rep}, in the thermodynamic limit this becomes
\begin{equation}
\label{qa-d-ensemble}
\quad\langle{\mathcal O}\rangle_d=\int{\mathcal D}\pmb{\rho}\exp\Big[ -2 {\cal E}(\pmb{\rho})+S_{YY}(\pmb{\rho})\Big]\langle\pmb{\rho}|{\mathcal O}|
\pmb{\rho}\rangle,
\end{equation}
where we defined the thermodynamic limit of the logarithm of the overlaps as \cite{foot2}
\begin{equation}
 {\cal E}(\pmb{\rho})\equiv-\lim_{\rm th} \Big[{\rm Re}(\ln\langle \pmb{\rho}|\Psi_0\rangle)\Big].
\end{equation}
This functional integral can be evaluated in the limit 
$L\to\infty$ using the saddle point approximation, since both the $S_{YY}(\pmb{\rho})$ and $ {\cal E}(\pmb{\rho})$
are generally estensive. 
One has to minimise the  functional appearing in the exponential in \eqref{qa-d-ensemble}, i.e.  
\begin{equation}
\label{qa-term}
{\mathcal S}_Q(\pmb{\rho})\equiv  -2 {\cal E}(\pmb{\rho})+S_{YY}(\pmb{\rho})  ,
\end{equation}
which has been termed {\it quench action}.
The minimisation is carried out solving 
\begin{equation}
\label{min}
\frac{ \delta{\mathcal S_Q}(\pmb{\rho})}{\delta\pmb{\rho}}\bigg|_{\pmb{\rho}=\pmb{\rho^*}}=0,
\end{equation}
under the constraint that the thermodynamic Bethe equations hold. 
Notice that  $S_{YY}$ appearing in \eqref{qa-d-ensemble} and \eqref{qa-term} 
counts only those microstates with a non zero-overlap with the initial state (i.e. those contributing to the sum \eqref{d-ensemble} before
taking the thermodynamic limit). 
This is only a small fraction of the total ones: in all quenches solved so far only parity invariant Bethe states 
(i.e. those containing only pairs of rapidities with opposite sign, i.e., such that $\{\lambda_j\}_{j=1}^M=\{-\lambda_j\}_{j=1}^M$) 
have non-zero overlap \cite{grd-10,cd-12,dwbc-14,pozsgay-14,bdwc-14,pc-14,fz-16,dkm-16,hst-17,msca-16}. 
This fact can be effectively taken into account either by performing the integral defining $S_{YY}$ \eqref{y-y} only on positive $\lambda$'s
or  multiplying $S_{YY}$ by $1/2$. We will adopt the latter choice.

Finally for the expectation values of observables~\eqref{qa-d-ensemble},  one obtains 
\begin{equation}
\label{obs-th}
\langle{\mathcal O}\rangle_d=\langle\pmb{\rho^*}|{\mathcal O}|\pmb{\rho^*}\rangle,
\end{equation}
under the reasonable assumption that the considered operator does not shift the saddle point (as obvious for local ones).
As a particular case, the identity operator in \eqref{d-ensemble} provides the normalisation of the initial state $\langle \Psi_0|\Psi_0\rangle=1$,
implying that at the saddle point the quench action vanishes, i.e.
\begin{equation}
{\cal S}_Q( \pmb{\rho^*})=0.
\label{norm}
\end{equation}

{\bf Overlap TBA for the diagonal entropies}.--
So far we have collected all the needed ingredients to study the diagonal entropy with the quench action. 
Taking the thermodynamic limit \eqref{rep} of the definition \eqref{d-renyi} with $\rho=\rho_d$ in \eqref{rho-diag}, 
we simply have 
\begin{equation}
\label{qa}
\textrm{Tr}\rho^\alpha_{d}=\int{\mathcal D}\pmb{\rho}e^{-2\alpha {\cal E}[\pmb{\rho}]+
\frac{1}{2}S_{YY}[\pmb{\rho}]},
\end{equation}
where we introduced the factor $1/2$ in front of $S_{YY}$ to explicitly take into account that only parity-invariant 
eigenstates have non-zero overlap with most initial states, as discussed above. 
The most important aspect of \eqref{qa} is that the R\'enyi index $\alpha$ appears in the exponential term 
and so it shifts the saddle points from the one in \eqref{min}. 

The modified quench action which should be considered in this case explicitly depends on the R\'enyi index as 
\begin{equation}
\label{f}
{\cal S}_{Q}^{(\alpha)}(\pmb{\rho})\equiv -2\alpha{\cal E}(\pmb{\rho})+\frac{1}{2}S_{YY}(\pmb{\rho}),
\end{equation}
and the quench action saddle-point equation is 
\begin{equation}
\label{mina}
\frac{ \delta{\mathcal S_Q^{(\alpha)}}(\pmb{\rho})}{\delta\pmb{\rho}}\bigg|_{\pmb{\rho}=\pmb{\rho^*_\alpha}}=0.
\end{equation}
The R\'enyi diagonal entropies are then the saddle point expectation of this quench action
\begin{equation}
\label{d-renyi-main}
S_d^{(\alpha)}=\frac{{\cal S}_Q^{(\alpha)}(\pmb{\rho}^*_\alpha)}{1-\alpha}=
\frac{1}{1-\alpha}\Big[-2\alpha{\cal E}(\pmb{\rho}^*_\alpha)+\frac{1}{2}S_{YY}(\pmb{\rho}^*_\alpha)\Big].
\end{equation}
${\cal S}_Q^{(\alpha)}(\pmb{\rho}^*_\alpha)$ can be thought of as a generalised free energy, with ${\cal E}$ 
playing the role of the Hamiltonian. 
Using the property that ${\cal S}_Q^{(1)}(\pmb{\rho}^*_1)=0$ as in \eqref{norm},
it is convenient to rewrite \eqref{d-renyi-main} as 
\begin{equation}
S_d^{(\alpha)}=\frac{{\cal S}_Q^{(\alpha)}(\pmb{\rho}^*_\alpha)- \alpha {\cal S}_Q^{(1)}(\pmb{\rho}^*_1)}{1-\alpha},
\label{s-rep}
\end{equation}
which is a form that closely resembles the replica approach to the entanglement entropy \cite{cc-04}. 

The  set of coupled equations \eqref{mina} can be solved, at least numerically, by standard methods. 
This analysis will be presented elsewhere \cite{ac-prep}, while here we only discuss some general properties of the 
diagonal R\'enyi entropies following from  \eqref{mina} and \eqref{d-renyi-main}.

%{\bf Limit $\alpha\to 0$}.-- For $\alpha\to 0$, the diagonal entropy  counts the number of eigenstates with non-zero overlap 
%with the initial state. When only parity invariant states have non-zero overlaps and consequently
%$S_d^{(0)}$ is half the Yang-Yang entropy of the thermal ensemble at infinite temperature. 

{\bf Proof that the diagonal entropy is half of the Yang-Yang entropy}.-- 
We now consider the limit 
\begin{equation}
\label{limit}
\lim_{\alpha\to 1}S^{(\alpha)}_d=S_d\equiv -{\rm Tr} \rho_d\ln\rho_d. 
\end{equation}
The normalisation \eqref{norm}, implying ${\cal S}_Q^{(1)}(\pmb{\rho}^*_1)=0$, ensures the limit to be finite. 
We then need to expand at the first order in $(\alpha-1)$ the saddle point quench action as
\begin{equation}
{\cal S}_Q^{(\alpha)} (\pmb{\rho}^*_\alpha)= \frac{d {\cal S}_Q^{(\alpha)} (\pmb{\rho}^*_\alpha)}{d\alpha}\bigg|_{\alpha=1} (\alpha-1) +O((\alpha-1)^2).
\end{equation}
We  have 
\begin{equation}
\frac{d {\cal S}_Q^{(\alpha)} (\pmb{\rho^*_\alpha})}{d\alpha}= 
\frac{\delta {\cal S}_Q^{(\alpha)} (\pmb{\rho})}{\delta \pmb{\rho}}\bigg|_{\pmb{\rho}= \pmb{\rho^*_\alpha}} 
\frac{\partial{\pmb{\rho^*_\alpha}}}{\partial \alpha}
+ \frac{\partial {\cal S}_Q^{(\alpha)} (\pmb{\rho})}{\partial \alpha}\bigg|_{\pmb{\rho}= \pmb{\rho^*_\alpha}},
\end{equation}
in which the first term in the rhs is zero because of the saddle-point condition \eqref{mina}
while the second term is simply $-2 {\cal E}(\pmb{\rho^*_\alpha})$ (cf. \eqref{f}). 
Hence we have
\begin{equation}
{\cal S}_Q^{(\alpha)} (\pmb{\rho}^*_\alpha)= -2 {\cal E}(\pmb{\rho}^*_1) (\alpha-1) +O((\alpha-1)^2),
\end{equation}
i.e. 
\begin{equation}
\lim_{\alpha\to 1}S^{(\alpha)}_d= -2 {\cal E}(\pmb{\rho}^*_1) = \frac12 S_{YY}(\pmb{\rho}^*_1) ,
\end{equation}
where in the last equality we used the normalisation conditions \eqref{norm}.
This proves that {\it the diagonal entropy is half of the thermodynamic one}. 
The sole assumption to derive this general result is that only parity invariant states have non-zero overlaps.
While this is a technical assumption, it is exactly on the 
 same line of thoughts of the physical interpretation of Gurarie \cite{gurarie-2013}. 
If for some  quenches (still unknown in integrable models) all states would have non-zero overlaps, 
then the above proof would lead to the equality between the diagonal and the thermodynamic entropy. 
If one imagine of generalising the quench action approach to non-integrable model (as speculated in \onlinecite{caux-16}) 
then the above argument would immediately lead to the equality of the two entropies, as numerically observed \cite{santos-2011}.

Indeed, our formalism can be used in general to show that R\'enyi diagonal entropies  are half of the 
thermodynamic R\'enyi entropies for arbitrary $\alpha$. Anyhow, the derivation is slightly more technical and so we report it in
the appendix. 
%Notice that the diagonal entropy $S_d$ is determined by the saddle point at $\alpha=1$. 
%This is a consequence of the fact that the logarithm in the definition of $S_d$ \eqref{limit} cannot shift the saddle point. 

{\bf Free fermionic systems}.--
For translational invariant systems of free fermions (or models mappable to), the diagonal R\'enyi entropies after a quench can be written 
in terms of the momentum occupation numbers $n_k$ as 
\begin{equation}
S^{(\alpha)}_d=\frac{L}{1-\alpha}\int \frac{dk}{2\pi} \ln [n_k^\alpha +(1-n_k)^\alpha],
\label{Sff}
\end{equation}
which simply tells that each mode is populated with probability $n_k$ and empty with probability $1-n_k$. 
Again, assuming parity invariance, the integral is only on positive $k$.
In TBA language, the rapidities $\lambda$ are the real momenta $k$; there is only one root density $\rho(k)=n_k/(2\pi)$ and
the hole density is  $\rho^{(h)}(k)=(1-n_k)/(2\pi)$. 

Thus, Eq. \eqref{Sff} implies that, for free fermions, the R\'enyi entropies for arbitrary $\alpha$ can be written in terms of the root 
density at $\alpha=1$ without the need of %writing and solving 
generalised TBA equations.
It is instructive to recover \eqref{Sff} from the generalised TBA saddle point \eqref{d-renyi-main} and show that, in this particular case, 
it is possible to write $\rho_\alpha$ in terms of $n_k, i.e. $ $\rho_{\alpha=1}$.
In the following we will work with $\tilde \rho(k)=2\pi \rho(k)$ to get rid of factors $2\pi$'s and often omit the $k$ dependence.

The Yang-Yang entropy for free fermions is 
\begin{equation}
\label{syy-ff}
S_{YY}=-L\int_{-\infty}^{\infty}\frac{dk}{2\pi}[\tilde \rho\ln \tilde \rho+(1-\tilde \rho)\ln(1-\tilde \rho)], 
\end{equation}
because $\rho^{(t)}=1/(2\pi)$. (For lattice free fermions, the integral runs between $-\pi$ and $\pi$, but this does not influence 
any of the following results.)   
Exploiting the parity invariance, the driving term ${\cal E}$ can be always written as 
\begin{equation}
{\cal E}=-L\Big[\int_0^\infty\frac{d k}{2\pi} \tilde \rho(k) \ln g(k)\Big], 
\end{equation}
for some calculable $g(k)$ depending on the initial state (e.g. for the quench from zero to infinite interaction in the Lieb-Liner gas 
$g(k)=4n^2/k^2$ \cite{grd-10} with $n$ being the density). 
The solution of the generalised quench action saddle point equation \eqref{mina} is straightforwardly worked out  
\begin{equation}
\tilde \rho^*_\alpha(k)= \frac1{1+[g(k)]^\alpha}\,.
\end{equation}
The mode occupation is  $\tilde \rho^*_1(k)$, i.e.  $n_k=1/(1+g(k))$.

In terms of $\tilde \rho^*_\alpha(k)$ the generalised  saddle-point free energy ${\cal S}_\alpha(\tilde \rho^*_\alpha)$ is, after simple algebra, 
\begin{equation}
{\cal S}_\alpha(\tilde \rho^*_\alpha)= - L\int_{0}^{\infty}\frac{dk}{2\pi}\ln (1-\tilde \rho^*_\alpha(k)).
\end{equation}
One can use the relation between $\tilde \rho^*_\alpha$ and $g(k)$ to 
write $\tilde \rho^*_\alpha$ as function of $n_k$ as
\begin{equation}
\tilde \rho^*_\alpha(k)= \frac{n_k^\alpha}{n_k^\alpha+(1-n_k)^\alpha}\,.
\end{equation}
The Renyi entropy is  straightforwardly obtained from  \eqref{s-rep} 
\begin{multline}
S^{(\alpha)}_d=-\frac{L}{1-\alpha}\int_{0}^{\infty}\frac{dk}{2\pi} \ln \frac{1-\tilde \rho^*_\alpha(k)}{(1-\tilde \rho^*_1(k))^\alpha}\\
 =\frac{L}{1-\alpha}\int_{0}^{\infty}\frac{dk}{2\pi}  \ln [{n_k^\alpha+(1-n_k)^\alpha}],
\end{multline}
which is  the same as \eqref{Sff}. 
Notice that to show this equality in a simple and general way we had to add to the saddle-point generalised action the 
term $\alpha S(\tilde \rho^*_1)$ which is zero, but written in terms of an integral of a non-zero integrand. % (i.e. $\ln (1-n_k)$).
%This  shows that there is an ambiguity in the form of the diagonal entropy as an integral over the rapidities. 
%Many general results derived by means of quench action here, such as the relation between mode occupation and overlaps, 
%agree with the explicit calculations in free fermionic model, like the interaction quench in the Ising model \cite{} and the quench 
%from zero to infinite interaction in the Lieb-Liniger model \cite{}. 

The same derivation above also holds for free bosons with very minor modifications. 

\begin{figure}[t]
\begin{center}
\includegraphics*[width=0.95\linewidth]{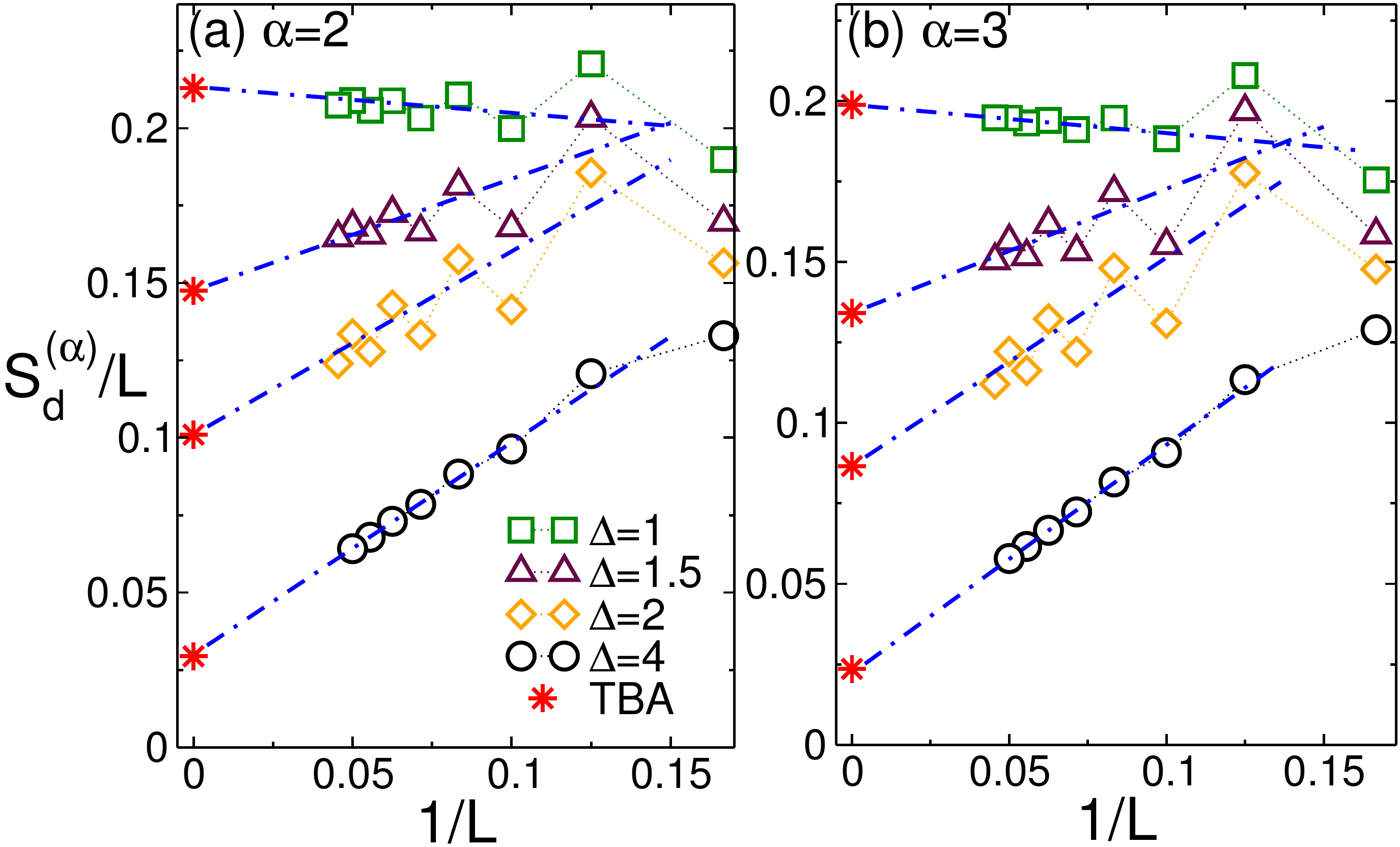}
\end{center}
\caption{ Diagonal R\'enyi entropies after the quench from  the N\'eel state in the $XXZ$ chain. 
 $S_{d}^{{}_{\textrm (\alpha\textit)}}/L$ is plotted against the inverse chain length $1/L$, for several values of the anisotropy parameter $\Delta$. 
 Panels (a) and (b) correspond to $\alpha=2$ and $\alpha=3$, respectively. 
 The stars are the TBA results in the thermodynamic limit. 
 The dash-dotted lines are finite size fits, as explained in the text. 
}
\label{diag-ed}
\end{figure}

{\bf A numerical example for an interacting model}.--
It is far beyond the aim of this paper to present detailed calculations of the R\'enyi diagonal entropies 
for interacting integrable models which require a large amount of technicalities. 
%This will be reported elsewhere \cite{ac-prep}.
%
However, it is instructive to show at least one example of the comparison between  the quench action approach and the 
numerical computation for finite systems. 
We focus on the quench in the spin-$1/2$ XXZ chain starting from the N\'eel state which has been studied intensively in the 
literature \cite{wdbf-14,ac-16qa,pvc-16}.
The exact diagonalisation data for the diagonal entropies are reported in Fig. \ref{diag-ed} showing
$S_d^{(\alpha)}$ for $\alpha=2$ and $\alpha=3$. % (panel (a) and (b) in the Figure) plotted versus $1/L$. 
%In both panels the different symbols correspond to different values of $\Delta$. 
Finite-size effects are visible for both $\alpha=2$ and $\alpha=3$.
The star symbols denote the diagonal entropy densities $s_{\infty}^{(\alpha)}$ in the thermodynamic limit,  
which are obtained using~\eqref{d-renyi-main}. 
The dash-dotted lines are linear fits to  $S_d^{(\alpha)}=s^{(\alpha)}_{\infty} + a_\alpha/L$ with $s^{(\alpha)}_{\infty}$ fixed 
to the TBA prediction. 
The data are clearly compatible with the quench action prediction for all values of $\Delta$ although 
large oscillations in $L$ affect the numerical data.

%%%%%%%%%%%%%%%%%%%%%%%%%%%%%%%%%%
%

%#############-- CONCLUSIONS --########################################
{\bf Conclusions}.-- 
In this paper we developed a Bethe ansatz framework to calculate the diagonal R\'enyi entropies after a quantum quench in integrable models. 
The diagonal R\'enyi entropies are generalised 
free energies evaluated over a thermodynamic macrostate which depends on the R\'enyi index $\alpha$ and
it is not the same describing the von Neumann entropy and the local observables.
We checked that this prescription reproduces known results for free models and we tested numerically its 
validity for interacting integrable models. 
An intriguing consequence is that the steady state contains information about different regions of the spectrum 
of the $XXZ$ chain, which can be accessed by varying $\alpha$. 
This is similar, in spirit, to the observation~\cite{garrison-2015} 
that a single eigenstate of a generic (non-integrable) Hamiltonian at 
finite energy density contains information about the full spectrum of the Hamiltonian. 
A byproduct of our approach is a very simple proof of the long standing issue that the 
diagonal entropy is half of the thermodynamic one in the steady state.
This is a direct consequence of the parity invariance of the Bethe eigenstates with non-zero overlap with the 
initial state.

This Letter opens several new questions. 
First, %from the technical point of view, 
it would be important to understand whether the R\'enyi entropies can be re-written in terms of the saddle point 
root densities describing local observables.
This does not look neither simple nor reasonable from our approach, but, 
%the explicit result for free fermions suggests that it could be possible. 
%
physically, the thermodynamic R\'enyi entropy is the same as the entanglement one for a large subsystem embedded in an infinite system. 
Thus it is, in principle, expected to be describable by the same saddle-point of the von Neumann entropy. 
Having the R\'enyi entropies written in terms 
of the standard root densities would allow to reconstruct the  entire time dependence 
of the entanglement R\'enyi entropies exploiting  the quasi-particle picture~\cite{calabrese-2005}, similarly to what  done for the von 
Neumann entropy~\cite{alba-2016}.  

Another issue is the role played by the parity invariance of the relevant post-quench eigenstates. 
Up to now this has been seen only as a technical remark (see however \onlinecite{delfino-14}), but 
we showed that it affects the ratio between  diagonal and thermodynamic entropies. 
%It would be interesting to understand whether it 
Is it possible to design a solvable quench where non-parity invariant 
states have non vanishing overlaps so that the ratio between the entropies equals one?
%, as for non integrable models \cite{santos-2011} 
(the arguments in \onlinecite{delfino-14} suggests this to be unlikely). 
Alternatively, are there quenches in which only a small subclass of parity invariant eigenstates have 
non-zero overlap  making the ratio between the two entropies even smaller than $1/2$? 

{\bf Acknowledgments}.--
We are grateful to Bruno Bertini for suggesting the proof reported in the appendix.
PC acknowledges support from the ERC under the Starting Grant No 279391 EDEQS and VA 
the European Union's Horizon 2020 under the Marie Sklodowoska-Curie grant agreement No 702612 OEMBS. 

{\bf Note added}.-- 
After the completion of this work, one of us and coworkers found a set of initial states for free-like theories in which 
the eigenstates with non-zero overlap are different from parity invariant states \cite{btc-17}. 
The analysis of the corresponding entropies is not trivial and still under examination.

\appendix 
\section{R\'enyi diagonal entropies are half of thermodynamic ones}
\label{app}

In this appendix we prove that  R\'enyi diagonal entropies are half of thermodynamic one for arbitrary index $\alpha$, generalizing the 
result for $\alpha=1$ presented in the main text.
However, this proof turn out to be more cumbersome and technical than the one for the von Neumann entropy,
mainly because the thermodynamic R\'enyi entropies must be derived from the GGE since there is no analogous of the Yang-Yang entropy. 
In particular, the proof requires the manipulation of the explicit TBA equations that we avoided in the main text. 

In TBA language, the saddle point equation (\ref{min}) is a set of coupled integral equations for the root and hole densities  
$\rho_n(\lambda)$ and $\rho^{(h)}_n(\lambda)$ for each string of length $n$.
These equations are written in a more compact form in terms of the ratios $\eta_n(\lambda)\equiv\rho^{(h)}_n(\lambda)/\rho_n(\lambda)$, 
assuming the standard form \cite{taka-book,caux-16}
\begin{equation}
\label{qa-tba}
\ln\eta_n(\lambda)=g_n(\lambda)+\sum\limits_{m=1}^\infty a_{nm}\star\ln (1+\eta_m^{-1})(\lambda), 
\end{equation}
where the symbol $\star$ denotes the convolution and
we introduced the driving term $g_n(\lambda)$ from
\begin{equation}
\label{driving}
{2\cal E}[\pmb\rho]=L\sum_{n=1}^\infty\int_0^\infty d\lambda\rho_n(\lambda) g_n(\lambda). 
\end{equation}
In Eq. (\ref{qa-tba}) the functions $a_{nm}(\lambda)$ encode the interaction in the thermodynamic limit and are simply read 
from the Bethe equations of the specific model (see e.g. \onlinecite{taka-book}).
Consequently the saddle point equation for general $\alpha$ in Eq. (\ref{mina}) are simply
\begin{equation}
\label{qa-tba-a}
\ln\eta_n(\lambda)=\alpha g_n(\lambda)+\sum\limits_{m=1}^\infty a_{nm}\star\ln (1+\eta_m^{-1})(\lambda).
\end{equation}

The generalized Gibbs ensemble describing local observables after a quench to an integrable model has the form 
\begin{equation}
\rho_{\rm GGE}=\exp\Big(-\sum_{k} \beta_k \hat I_k\Big) , 
\end{equation}
where the operators $\hat I_k$'s form a complete set of local and quasi-local integrals of motion and $\beta_k$ are fixed by the condition that 
the charges assume the same value in the initial state and in the GGE, i.e. 
$\langle \Psi_0| \hat I_k| \Psi_0\rangle= {\rm Tr} [\rho_{\rm GGE} \hat I_k]$.

The expectation values of local operator in the GGE are obtained by generalized TBA \cite{mc-12}.
Analogously to (\ref{qa-d-ensemble}) for the diagonal ensemble, we write the expectation value of a local operator $O$ as a path integral 
 \begin{equation}
\langle O\rangle_{\rm GGE}
=\int{\mathcal D}\pmb{\rho}\exp\Big[ -\sum_k \beta_k I_k(\pmb{\rho})+S_{YY}(\pmb{\rho})\Big]\langle\pmb{\rho}|{\mathcal O}|
\pmb{\rho}\rangle,
\end{equation}
where we used the fact that Bethe states $|\pmb \rho\rangle$ are eigenstates of the conserved charges with eigenvalues $I_k(\pmb{\rho})$, i.e. 
\begin{equation}
\hat I_k |\pmb{\rho}\rangle=  I_k(\pmb{\rho}) |\pmb{\rho} \rangle. 
\end{equation}
The resulting saddle point equations are easily read from the above path integral \cite{mc-12}
\begin{equation}
\label{qa-gge}
\ln\eta_n(\lambda)=\sum_{k} \beta_k f_{kn}(\lambda)+\sum\limits_{m=1}^\infty a_{nm}\star\ln (1+\eta_m^{-1})(\lambda), 
\end{equation}
where we introduced the functions $f_{kn}(\lambda)$ as 
\begin{equation}
 I_k(\pmb \rho)= L \sum_n \int_{-\infty}^\infty d\lambda f_{kn} (\lambda)\rho_n(\lambda).
 \label{Ivslam}
\end{equation}
These functions $f_{kn}(\lambda)$ are the representation of the local and quasi-local integrals of motion in terms of the rapidities and are 
generically known but can be cumbersome. For example in the case of the repulsive Lieb-Liniger model where there are no strings
(i.e. there is not the sum over $n$ above), these functions are simply $f_k(\lambda)=\lambda^k$, i.e.  
$ I_k(\pmb \rho)= L \int d\lambda \lambda^k \rho(\lambda)$.

At this point, since the expectation values of all the local observables in the diagonal ensemble and in the GGE must be equal, 
these must have the same saddle point root distribution $\pmb \rho$ which implies Eqs. (\ref{qa-tba}) and (\ref{qa-gge}) to be equal. 
This equality leads to the identification 
\begin{equation}
 \sum_k  \beta_k f_{kn}(\lambda)=g_n(\lambda)\,,
\label{id1}
\end{equation}
for $\lambda\geq0$ where $g_n(\lambda)$ is defined. 
This identification implies 
\begin{equation}
4 {\cal E}(\pmb\rho)= \sum_k \beta_k I_k (\pmb\rho)\,.
\label{id2}
\end{equation}
where we used the fact that the integral in (\ref{driving}) is on the positive axis, while the one in (\ref{Ivslam}) is on the entire real line. 
Notice that those functions $f_{kn}(\lambda)$ which are odd in $\lambda$ do not contribute to the GGE because the considered initial state 
is even and the corresponding $\beta_k$'s are vanishing. This consideration ensures the validity of (\ref{id2}).

We are finally ready to obtain thermodynamic R\'enyi entropies. 
These are given by 
\begin{equation}
\textrm{Tr}\rho^\alpha_{\rm GGE}=\int{\mathcal D}\pmb{\rho}e^{-\alpha \sum_k \beta_k I_k[\pmb \rho]+ S_{YY}[\pmb{\rho}]}.
\end{equation}
Thus, the saddle point equation are the same as in (\ref{qa-gge}) with the replacement $\beta_k\to \alpha \beta_k$ for all $k$.
Because of the identification (\ref{id1}), these generalised TBA for the GGE equations are the same as those for the diagonal 
ensemble (\ref{qa-tba-a}). This equivalence could have been expected, but it is far from trivial because R\'enyi entropies are global and 
not local quantities (and indeed we will soon find a factor 2 between the two). 
At this point, the GGE R\'enyi entropy is read off from the saddle-point of the above path integral, in analogy to the 
diagonal one (\ref{d-renyi-main}), as
\begin{equation}
S_{\rm GGE}^{(\alpha)}=
\frac{1}{1-\alpha}\Big[-\alpha \sum_k \beta_k I_k(\pmb{\rho}^*_\alpha)+S_{YY}(\pmb{\rho}^*_\alpha)\Big].
\end{equation}
which using (\ref{id2}) is exactly twice the diagonal R\'enyi entropy (\ref{d-renyi-main}).

%#############-- BIBLIOGRAPHY --########################################

%


\begin{thebibliography}{99}


\section*{References}

\bibitem{v-29}
J.~von Neumann, Beweis des Ergodensatzes und des H-Theorems, 
\href{http://dx.doi.org/10.1007/BF01339852}{Z Phys. {\bf 57}, 30 (1929)}.

\bibitem{eth0}
R.~V.~Jensen and R.~Shankar, Statistical Behavior in Deterministic Quantum Systems with Few Degrees of Freedom, 
\href{http://dx.doi.org/10.1103/PhysRevLett.54.1879}{Phys. Rev. Lett. {\bf 54}, 1879 (1985).}

\bibitem{deutsch-1991}
J.~M.~Deutsch, Quantum statistical mechanics in a closed system, 
\href{http://dx.doi.org/10.1103/PhysRevA.43.2046}{Phys. Rev. A {\bf 43}, 2046 (1991)}..

\bibitem{srednicki-1994}
M.~Srednicki, Chaos and quantum thermalization,  \href{http://dx.doi.org/10.1103/PhysRevE.50.888}{Phys. Rev. E {\bf 50}, 888 (1994)}. 

\bibitem{rdo-08}  
M. Rigol, V. Dunjko, and M. Olshanii, Thermalization and its mechanism for generic isolated quantum systems, 
\href{http://dx.doi.org/10.1038/nature06838}{Nature {\bf 452}, 854 (2008)}.

\bibitem{rigol-2012}
M. Rigol  and M. Srednicki, Alternatives to Eigenstate Thermalization,  
\href{http://dx.doi.org/10.1103/PhysRevLett.100.100601}{Phys. Rev. Lett. {\bf 108}, 110601 (2012)}..

\bibitem{dalessio-2015}
L.~D'Alessio, Y.~Kafri, A.~Polkovnikov, and M.~Rigol,  
From Quantum Chaos and Eigenstate Thermalization to Statistical Mechanics and Thermodynamics, 
\href{http://dx.doi.org/10.1080/00018732.2016.1198134}{Adv.\ Phys.\ {\bf 65}, 239 (2016)}. 

\bibitem{rigol-2007}
M. Rigol, V. Dunjko, V. Yurovsky, and M. Olshanii, Relaxation in a Completely 
Integrable Many-Body Quantum System: An {\it Ab Initio} Study of the Dynamics of 
the Highly Excited States of 1D Lattice Hard-Core Bosons. 
\href{http://dx.doi.org/10.1103/PhysRevLett.98.050405}{Phys. Rev. Lett. {\bf 98}, 050405 (2007)}. 

\bibitem{cazalilla-2006}
M.~A.~Cazalilla, 
Effect of Suddenly Turning on Interactions in the Luttinger Model, 
\href{http://dx.doi.org/10.1103/PhysRevLett.97.156403}{Phys. Rev. Lett. {\bf 97}, 156403 (2006)}.

\bibitem{barthel-2008}
T.~Barthel and U.~Schollw\"ock,  
Dephasing and the Steady State in Quantum Many-Particle Systems. 
\href{http://dx.doi.org/10.1103/PhysRevLett.100.100601}{Phys. Rev. Lett. {\bf 100}, 100601 (2008)}.

\bibitem{cramer-2008}
M.~Cramer, C.~M.~Dawson, J.~Eisert, and T.~J.~Osborne,  
Exact Relaxation in a Class of Nonequilibrium Quantum Lattice Systems, 
\href{http://dx.doi.org/10.1103/PhysRevLett.100.030602}{Phys. Rev. Lett. {\bf 100}, 030602 (2008)};\\
M.~Cramer and J.~Eisert,  
A quantum central limit theorem for non-equilibrium systems: exact 
local relaxation of correlated states, 
\href{http://dx.doi.org/10.1088/1367-2630/12/5/055020}{New\ J.\ Phys.\ {\bf 12}, 055020 (2010)}.

\bibitem{fagotti-2008}
M.~Fagotti and P.~Calabrese,  
Evolution of entanglement entropy following a quantum quench: Analytic results for the XY chain in a transverse magnetic field. 
\href{https://doi.org/10.1103/PhysRevA.78.010306}{Phys. Rev. A 78, 010306 (2008)}. 

\bibitem{cazalilla-2012a}
M.~A.~Cazalilla, A.~Iucci, and M.-C.~Chung, 
Thermalization and quantum correlations in exactly solvable models, 
\href{http://dx.doi.org/10.1103/PhysRevE.85.011133}{Phys. Rev. E {\bf 85}, 011133 (2012)}.

\bibitem{calabrese-2012}
P. Calabrese, F. H. L. Essler, and M. Fagotti, Quantum Quench in the Transverse-Field Ising Chain,
\href{http://dx.doi.org/10.1103/PhysRevLett.106.227203}{Phys. Rev. Lett. {\bf 106}, 227203 (2011)}; \\
P. Calabrese, F. H. L. Essler, and M. Fagotti, Quantum quench in the transverse field Ising chain: I. Time evolution of order parameter correlators,
\href{http://dx.doi.org/10.1088/1742-5468/2012/07/P07016}{J. Stat. Mech. (2012) P07016}; \\
P. Calabrese, F. H. L. Essler, and M. Fagotti, Quantum quenches in the transverse field Ising chain: II. Stationary state properties,
\href{http://dx.doi.org/10.1088/1742-5468/2012/07/P07022}{J. Stat. Mech. (2012) P07022}.


\bibitem{sotiriadis-2012}
D. Fioretto and G. Mussardo, Quantum quenches in integrable field theories,
\href{http://dx.doi.org/10.1088/1367-2630/12/5/055015}{New J. Phys. {\bf 12}, 055015 (2010)};\\
S. Sotiriadis, D. Fioretto, and G. Mussardo, Zamolodchikov-Faddeev algebra and quantum quenches in integrable field theories,
\href{http://dx.doi.org/10.1088/1742-5468/2012/02/P02017}{J. Stat. Mech. (2012) P02017}.

\bibitem{collura-2013}
M. Collura, S. Sotiriadis, and P. Calabrese, Equilibration of a Tonks-Girardeau Gas Following a Trap Release,
\href{http://dx.doi.org/10.1103/PhysRevLett.110.245301}{Phys. Rev. Lett. {\bf 110}, 245301 (2013)};\\
M. Collura, S. Sotiriadis, and P. Calabrese, Quench dynamics of a Tonks-Girardeau gas released from a harmonic trap,
\href{http://dx.doi.org/10.1088/1742-5468/2013/09/P09025}{J. Stat. Mech. (2013) P09025}.

\bibitem{fagotti-2013}
M.~Fagotti and F.~H.~L.~Essler, Stationary behaviour of observables after a quantum quench in the spin-$1/2$ Heisenberg $XXZ$ chain, 
\href{http://dx.doi.org/10.1088/1742-5468/2013/07/P07012}{J. Stat. Mech. (2013) P07012}.

\bibitem{p-13}
B. Pozsgay, The generalized Gibbs ensemble for Heisenberg spin chains,
\href{http://dx.doi.org/10.1088/1742-5468/2013/07/P07003}{J. Stat. Mech. (2013) P07003}.

\bibitem{fe-13b}
M. Fagotti and F. H. L. Essler, Reduced Density Matrix after a Quantum Quench,
\href{http://dx.doi.org/10.1103/PhysRevB.87.245107}{Phys. Rev. B {\bf 87}, 245107 (2013)}.

\bibitem{kcc14}
M.~Kormos, M.~Collura, and P.~Calabrese,  
Analytic results for a quantum quench from free to hard-core one-dimensional bosons, 
\href{http://dx.doi.org/10.1103/PhysRevA.89.013609}{Phys. Rev. A {\bf 89}, 013609 (2014)}.

\bibitem{sotiriadis-2014}
S.~Sotiriadis and P.~Calabrese,  
Validity of the GGE for quantum quenches from interacting to noninteracting models, 
\href{http://dx.doi.org/10.1088/1742-5468/2014/07/P07024}{J. Stat. Mech. (2014) P07024}. 

\bibitem{fcec-14}
M. Fagotti, M. Collura, F. H. L. Essler, and P. Calabrese, Relaxation after quantum quenches in the spin-12 Heisenberg XXZ chain,
\href{http://dx.doi.org/10.1103/PhysRevB.89.125101}{Phys. Rev. B {\bf 89}, 125101 (2014)}.

\bibitem{ilievski-2015a}
E.~Ilieveski, J.~De~Nardis, B.~Wouters, J.-S.~Caux, F.~H.~L.~Essler, and T.~Prosen,  
Complete Generalized Gibbs Ensembles in an Interacting Theory, 
\href{http://dx.doi.org/10.1103/PhysRevLett.115.157201}{Phys. Rev. Lett. {\bf 115}, 157201 (2015)};\\
E. Ilievski, E. Quinn, J. D. Nardis, and M. Brockmann, String-charge duality in integrable lattice models,
\href{http://dx.doi.org/10.1088/1742-5468/2016/06/063101}{J. Stat. Mech. (2016) 063101}.

\bibitem{alba-2015}
V.~Alba, Simulating the Generalized Gibbs Ensemble (GGE): a Hilbert space Monte Carlo approach. 
\href{http://arxiv.org/abs/1507.06994}{arXiv:1507.06994}. 

\bibitem{langen-15} 
T. Langen, S. Erne, R. Geiger, B. Rauer, T. Schweigler, M. Kuhnert, W. Rohringer, I. E. Mazets, T. Gasenzer, 
and J. Schmiedmayer, 
Experimental observation of a generalized Gibbs ensemble,
\href{http://dx.doi.org/10.1126/science.1257026}{Science {\bf 348}, 207 (2015)}.

\bibitem{essler-2015}
F. H. L. Essler, G. Mussardo, and M. Panfil, Generalized Gibbs ensembles for quantum field theories,
\href{http://dx.doi.org/10.1103/PhysRevA.91.051602}{Phys. Rev. A {\bf 91}, 051602 (2015)};\\
F. H. L. Essler, G. Mussardo, and M. Panfil, On Truncated Generalized Gibbs Ensembles in the Ising Field Theory,
\href{https://doi.org/10.1088/1742-5468/aa53f4}{J. Stat. Mech. (2017) 013103}.

\bibitem{cardy-2015}
J.~Cardy, Quantum quenches to a critical point in one dimension: some further results, 
\href{http://dx.doi.org/10.1088/1742-5468/2016/02/023103}{J. Stat. Mech. (2016) 023103}.

\bibitem{sotiriadis-2016}
S.~Sotiriadis,  Memory-preserving equilibration after a quantum quench in a 1d critical model,  
\href{https://doi.org/10.1103/PhysRevA.94.031605}{Phys. Rev. A {\bf 94}, 031605 (2016)}. 


\bibitem{bastianello-2016}
A.~Bastianello and S.~Sotiriadis, Quasi locality of the GGE in interacting-to-free  quenches in relativistic field theories, 
\href{https://doi.org/10.1088/1742-5468/aa5738}{J. Stat. Mech. (2017) 023105}.. 

\bibitem{vernier-2016}
E. Vernier and A. Cort\'es Cubero, Quasilocal charges and the complete GGE for field theories with non 
diagonal scattering, \href{http://dx.doi.org/10.1088/1742-5468/aa5288}{J. Stat. Mech. (2017) 23101}.. 

\bibitem{pvw-17} 
B. Pozsgay, E. Vernier, and M. A. Werner, On Generalized Gibbs Ensembles with an infinite set of conserved charges,
\href{http://arxiv.org/abs/1703.09516}{arXiv:1703.09516 (2017)}.

\bibitem{vidmar-2016}
L.~Vidmar and M.~Rigol, Generalized Gibbs ensemble in integrable lattice models, 
\href{http://dx.doi.org/10.1088/1742-5468/2016/06/064007}{J. Stat. Mech. (2016) 064007}.

\bibitem{gogolin-2015}
C.~Gogolin and J.~Eisert, 
Equilibration, thermalisation, and the emergence of statistical mechanics in closed quantum systems, 
\href{http://iopscience.iop.org/article/10.1088/0034-4885/79/5/056001}{Rep. Prog. Phys. {\bf 79}, 056001 (2016)}. 

\bibitem{calabrese-2016}
P.~Calabrese, F.~H.~L.~Essler, and G.~Mussardo,  Introduction to ``Quantum Integrability in Out of Equilibrium Systems'', 
\href{http://dx.doi.org/10.1088/1742-5468/2016/06/064001}{J.\ Stat.\ Mech.\ (2016) P064001}. 



\bibitem{ef-16}
F. H. L. Essler and M. Fagotti, Quench dynamics and relaxation in isolated integrable quantum spin chains,
\href{http://dx.doi.org/10.1088/1742-5468/2016/06/064002}{J. Stat. Mech. (2016) 064002}.


\bibitem{foot1}
Hereafter, when stating that two entropies are the same, we mean that they have the same extensive behaviour (i.e. the same density), 
but they can have different subleading terms (and in most cases they do).



%%%%%%%%%%%%%



\bibitem{calabrese-2005}
P.~Calabrese and J.~Cardy, Evolution of Entanglement Entropy in One-Dimensional Systems, 
 \href{http://dx.doi.org/10.1088/1742-5468/2005/04/P04010}{J. Stat. Mech. (2005) P04010}.

\bibitem{dls-13}
J. M. Deutsch, H. Li, and A. Sharma, Microscopic origin of thermodynamic entropy in isolated systems, 
\href{http://dx.doi.org/10.1103/PhysRevE.87.042135}{Phys. Rev. E {\bf 87}, 042135 (2013)}.

\bibitem{spr-12}
L. F. Santos, A. Polkovnikov, and M. Rigol, Weak and strong typicality in quantum systems, 
\href{http://dx.doi.org/10.1103/PhysRevE.86.010102}{Phys. Rev. E {\bf 86}, 010102 (2012)}.



\bibitem{collura-2014}
M.~Collura, M.~Kormos, and P.~Calabrese, Stationary entropies following an interaction quench in $1D$ Bose gas, 
\href{https://doi.org/10.1088/1742-5468/2014/01/P01009}{J. Stat. Mech. P01009 (2014)}.

\bibitem{kormos-2014}
M.~Kormos, L.~Bucciantini, and P.~Calabrese, Stationary entropies after a quench from excited states in the Ising chain, 
\href{https://doi.org/10.1209/0295-5075/107/40002}{EPL 107, 40002 (2014)};\\
L. Bucciantini, M. Kormos, and P. Calabrese, Quantum quenches from excited states in the Ising chain, 
\href{https://doi.org/10.1088/1751-8113/47/17/175002}{J. Phys. A {\bf 47}, 175002 (2014)}.


\bibitem{bam-15}
W. Beugeling, A. Andreanov, and M. Haque,
Global characteristics of all eigenstates of local many-body Hamiltonians: participation ratio and entanglement entropy,
\href{https://doi.org/10.1088/1742-5468/2015/02/P02002}{J. Stat. Mech. (2015) P02002}.


\bibitem{kaufman-2016}
A.~M.~Kaufman, M.~E.~Tai, A.~Lukin, M.~Rispoli, R.~Schittko, P.~M.~Preiss, and  M.~Greiner, 
Quantum thermalization through entanglement in an isolated many-body system, 
\href{http://dx.doi.org/10.1126/science.aaf6725}{Science {\bf 353}, 794  (2016)}.

\bibitem{cc-16}
P. Calabrese and J. Cardy, 
Quantum quenches in 1+1 dimensional conformal field theories,
\href{http://iopscience.iop.org/article/10.1088/1742-5468/2016/06/064003}{J. Stat. Mech. (2016) 064003}.

\bibitem{alba-2016}
V.~Alba and P.~Calabrese, Entanglement and thermodynamics after a quantum quench in integrable 
systems, \href{http://dx.doi.org/10.1073/pnas.1703516114}{PNAS {\bf 114}, 7947 (2017)}. 

\bibitem{d-17}
J. Dubail, 
Entanglement scaling of operators: a conformal field theory approach, with a glimpse of simulability of long-time dynamics in 1+1d,
\href{http://dx.doi.org/10.1088/1751-8121/aa6f38}{J. Phys. A {\bf 50}, 234001 (2017)}.





\bibitem{polkovnikov-2011}
A.~Polkovnikov, Microscopic diagonal entropy and its connection to basic thermodynamic relations, 
\href{http://dx.doi.org/10.1016/j.aop.2010.08.004}{Ann.\ Phys.\ {\bf 326}, 486 (2011)}.


\bibitem{santos-2011}
L.~F.~Santos, A.~Polkovnikov, and M.~Rigol, Entropy of isolated quantum systems after a quench, 
\href{http://dx.doi.org/10.1103/PhysRevLett.107.040601}{Phys.\ Rev.\ Lett.\ {\bf 107}, 040601 (2011)}. 

\bibitem{gurarie-2013}
V.~Gurarie, Global large time dynamics and the generalized Gibbs ensemble, 
\href{https://doi.org/10.1088/1742-5468/2013/02/P02014}{J.\ Stat.\ Mech.\ (2013) P02014}. 

\bibitem{fagotti-2013b}
M. Fagotti, Finite-size corrections vs. relaxation after a sudden quench,
\href{http://dx.doi.org/10.1103/PhysRevB.87.165106}{Phys. Rev. B 87, 165106 (2013)}. 


\bibitem{dora-2014}
B.~D\'ora, Escort distribution function of work done and diagonal entropies in quenched Luttinger liquids, 
\href{http://dx.doi.org/10.1103/PhysRevB.90.245132}{Phys.\ Rev.\ B\ {\bf 90}, 245132 (2014)}. 

\bibitem{piroli-2016}
L.~Piroli, E.~Vernier, P.~Calabrese, and M.~Rigol, Correlations and diagonal entropies after quantum quenches in $XXZ$ chains, 
\href{http://dx.doi.org/10.1103/PhysRevB.95.054308}{Phys. Rev. B {\bf 95}, 054308 (2017)}.

\bibitem{v-00}
G. Vidal, Entanglement monotones, \href{http://dx.doi.org/10.1080/09500340008244048}{J. Mod. Opt. {\bf 47}, 355 (2000)}.

\bibitem{cc-04}
P. Calabrese and J. Cardy, Entanglement entropy and quantum field theory, 
\href{http://dx.doi.org/10.1088/1742-5468/2004/06/P06002}{J.  Stat. Mech. P06002 (2004)};\\
P. Calabrese and J. Cardy,  Entanglement entropy and conformal field theory, 
\href{http://dx.doi.org/10.1088/1751-8113/42/50/504005}{J. Phys. A {\bf 42}, 504005 (2009)}.

\bibitem{islam-2015}
R.~Islam, R.~Ma, P.~M.~Preiss, M.~E.~Tai, A.~Lukin, M.~Rispoli, and M.~Greiner,  
Measuring entanglement entropy in a quantum many-body system, 
\href{http://dx.doi.org/10.1038/nature15750}{Nature {\bf 528}, 77 (2015)}. 

\bibitem{caux-2013}
J.-S.~Caux and F.~H.~L.~Essler, Time evolution of local observables after quenching to an integrable model, 
\href{http://dx.doi.org/10.1103/PhysRevLett.110.257203}{Phys. Rev. Lett. {\bf 110}, 257203 (2013)}.

\bibitem{caux-16}
J.-S. Caux, The Quench Action,
\href{http://dx.doi.org/10.1088/1742-5468/2016/06/064006}{J. Stat. Mech. (2016) 064006}.



\bibitem{ac-prep}
V. Alba and P. Calabrese, in preparation. 

\bibitem{foot2}
The thermodynamic limit $\displaystyle \lim_{\rm th}$ is taken with the number of particles $N$ and the length $L$ going to infinity at fixed density $N/L$.

\bibitem{yy-69}
C. N. Yang and C. P. Yang, Thermodynamics of a One Dimensional System of Bosons with Repulsive Delta Function Interaction, 
\href{http://dx.doi.org/10.1063/1.1664947}{J. Math. Phys. {\bf 10}, 1115 (1969)}.

\bibitem{taka-book}
M.~Takahashi, {\it Thermodynamics of one-dimensional solvable models},
Cambridge University Press, Cambridge, 1999. 

\bibitem{dc-14}
J. De Nardis and J.-S. Caux, 
Analytical expression for a post-quench time evolution of the one-body density matrix of one-dimensional hard-core bosons,
\href{http://dx.doi.org/10.1088/1742-5468/2014/12/P12012}{J. Stat. Mech. (2014), P12012}.

\bibitem{dlc-15}
J. De Nardis, L. Piroli, and J.-S. Caux, Relaxation dynamics of local observables in integrable systems,
\href{http://dx.doi.org/10.1088/1751-8113/48/43/43FT01}{J. Phys. A {\bf 48}, 43FT01 (2015)}.

\bibitem{pc-17}
L. Piroli and P. Calabrese, Exact dynamics following an interaction quench in a one-dimensional anyonic gas,
\href{http://dx.doi.org/10.1103/PhysRevA.96.023611}{Phys. Rev. A {\bf 96}, 023611 (2017)}.

%%%%%%%%QA

\bibitem{BeSE14} 
B. Bertini, D. Schuricht, and F. H. L. Essler, Quantum quench in the sine-Gordon model,
\href{http://dx.doi.org/10.1088/1742-5468/2014/10/P10035}{J. Stat. Mech. (2014) P10035}.

\bibitem{BePC16} 
B. Bertini, L. Piroli, and P. Calabrese, Quantum quenches in the sinh-Gordon model: steady state and one-point correlation functions,
\href{http://dx.doi.org/10.1088/1742-5468/2016/06/063102}{J. Stat. Mech. (2016) 063102}.

\bibitem{pozsgay-13}
B. Pozsgay, The dynamical free energy and the Loschmidt echo for a class of quantum quenches in the Heisenberg spin chain,
\href{http://dx.doi.org/10.1088/1742-5468/2013/10/P10028}{J. Stat. Mech. (2013) P10028}.

\bibitem{wdbf-14} 
B. Wouters, J. De Nardis, M. Brockmann, D. Fioretto, M. Rigol, and J.-S. Caux, 
Quenching the Anisotropic Heisenberg Chain: Exact Solution and Generalized Gibbs Ensemble Predictions,
\href{http://dx.doi.org/10.1103/PhysRevLett.113.117202}{Phys. Rev. Lett. {\bf 113}, 117202 (2014)};\\
M. Brockmann, B. Wouters, D. Fioretto, J. D. Nardis, R. Vlijm, and J.-S. Caux, 
Quench action approach for releasing the N\'eel state into the spin-1/2 XXZ chain,
\href{http://dx.doi.org/10.1088/1742-5468/2014/12/P12009}{J. Stat. Mech. (2014) P12009}.

\bibitem{PMWK14} 
B. Pozsgay, M. Mesty\'an, M. A. Werner, M. Kormos, G. Zar\'and, and G. Tak\'acs, 
Correlations after Quantum Quenches in the $XXZ$ Spin Chain: Failure of the Generalized Gibbs Ensemble,
\href{http://dx.doi.org/10.1103/PhysRevLett.113.117203}{Phys. Rev. Lett. {\bf 113}, 117203 (2014)};\\
M. Mesty\'an, B. Pozsgay, G. Tak\'acs, and M. A. Werner, 
Quenching the XXZ spin chain: quench action approach versus generalized Gibbs ensemble,
\href{http://dx.doi.org/10.1088/1742-5468/2015/04/P04001}{J. Stat. Mech. (2015) P04001}.


\bibitem{ac-16qa} V.~Alba, and P.~Calabrese, The quench action approach in finite integrable spin chains, 
\href{http://dx.doi.org/10.1088/1742-5468/2016/04/043105}{J. Stat. Mech. (2016), 043105}.

\bibitem{ppv-17}
L. Piroli, B. Pozsgay, and E. Vernier, From the quantum transfer matrix to the quench action: the Loschmidt echo in XXZ Heisenberg spin chains,
\href{http://dx.doi.org/10.1088/1742-5468/aa5d1e}{J. Stat. Mech. (2017) 23106}.

\bibitem{mbpc-17}
M. Mestyan, B. Bertini, L. Piroli, and P. Calabrese, Exact solution for the quench dynamics of a nested integrable system,
\href{http://dx.doi.org/10.1088/1742-5468/aa7df0}{J. Stat. Mech. (2017) 083103}.


\bibitem{pce-16}
L. Piroli, P. Calabrese, and F. H. L. Essler, 
Multiparticle Bound-State Formation following a Quantum Quench to the One-Dimensional Bose Gas with Attractive Interactions,
\href{http://dx.doi.org/10.1103/PhysRevLett.116.070408}{Phys. Rev. Lett. {\bf 116}, 070408 (2016)};\\
L. Piroli, P. Calabrese, and F. H. L. Essler, Quantum quenches to the attractive one-dimensional Bose gas: exact results,
\href{http://dx.doi.org/10.21468/SciPostPhys.1.1.001}{SciPost Phys. {\bf 1}, 001 (2016)}.

\bibitem{Bucc16} 
L. Bucciantini, Stationary State After a Quench to the Lieb-Liniger from Rotating BECs,
\href{http://dx.doi.org/10.1007/s10955-016-1535-7}{J Stat Phys {\bf 164}, 621 (2016)}.

\bibitem{npg-17}
J. De Nardis, M. Panfil, A. Gambassi, L. F. Cugliandolo, R. Konik, and L. Foini,
Probing non-thermal density fluctuations in the one-dimensional Bose gas,
\href{http://arxiv.org/abs/1704.06649}{arXiv:1704.06649};\\
J. De Nardis and M. Panfil, Exact correlations in the Lieb-Liniger model and detailed balance out-of-equilibrium,
\href{http://dx.doi.org10.21468/SciPostPhys.1.2.015}{SciPost Phys. 1, 015 (2016)}.


\bibitem{dwbc-14}
J. De Nardis, B. Wouters, M. Brockmann, and J.-S. Caux, Solution for an interaction quench in the Lieb-Liniger Bose gas,
\href{http://dx.doi.org/10.1103/PhysRevA.89.033601}{Phys. Rev. A {\bf 89}, 033601 (2014)}.

%%%%%%%overlaps

\bibitem{grd-10}
V. Gritsev, T. Rostunov, and E. Demler, Exact methods in the analysis of the non-equilibrium dynamics of integrable models: application to the study of correlation functions for non-equilibrium 1D Bose gas,
\href{http://dx.doi.org/10.1088/1742-5468/2010/05/P05012}{J. Stat. Mech. (2010) P05012}.

\bibitem{cd-12}
P. Le Doussal and P. Calabrese,
The KPZ equation with flat initial condition and the directed polymer with one free end
\href{http://dx.doi.org/10.1088/1742-5468/2012/06/P06001}{J. Stat. Mech. (2012) P06001};\\
P. Calabrese and P. Le Doussal, 
Interaction quench in a Lieb-Liniger model and the KPZ equation with flat initial conditions
\href{http://dx.doi.org/10.1088/1742-5468/2014/05/P05004}{J. Stat. Mech. (2014) P05004}.


\bibitem{pozsgay-14}
B. Pozsgay, Overlaps between eigenstates of the XXZ spin-1/2 chain and a class of simple product states,
\href{http://dx.doi.org/10.1088/1742-5468/2014/06/P06011}{J. Stat. Mech. (2014) P06011}.

\bibitem{bdwc-14}
M. Brockmann, J. D. Nardis, B. Wouters, and J.-S. Caux, A Gaudin-like determinant for overlaps of N\'eel and XXZ Bethe states,
\href{http://dx.doi.org/10.1088/1751-8113/47/14/145003}{J. Phys. A {\bf 47}, 145003 (2014)};\\
M. Brockmann, Overlaps of q-raised N\'eel states with XXZ Bethe states and their relation to the Lieb-Liniger Bose gas,
\href{http://dx.doi.org/10.1088/1742-5468/2014/05/P05006}{J. Stat. Mech. (2014) P05006};\\
M. Brockmann, J. De Nardis, B. Wouters, and J.-S. Caux, N\'eel-XXZ state overlaps: odd particle numbers and Lieb-Liniger scaling limit,
\href{http://dx.doi.org/10.1088/1751-8113/47/34/345003}{J. Phys. A {\bf 47}, 345003 (2014)}.

\bibitem{pc-14}
L. Piroli and P. Calabrese, Recursive formulas for the overlaps between Bethe states and product states in XXZ Heisenberg chains,
\href{http://dx.doi.org/10.1088/1751-8113/47/38/385003}{J. Phys. A {\bf 47}, 385003 (2014)}.

\bibitem{fz-16}
M. de Leeuw, C. Kristjansen, and K. Zarembo,
One-point Functions in Defect CFT and Integrability,
\href{http://dx.doi.org/10.1007/JHEP08(2015)098}{JHEP 08 (2015) 098};\\
I. Buhl-Mortensen, M. de Leeuw, C. Kristjansen, and K. Zarembo,
One-point Functions in AdS/dCFT from Matrix Product States,
\href{http://dx.doi.org/10.1007/JHEP02(2016)052}{JHEP 02 (2016) 052};\\
O. Foda and K. Zarembo, Overlaps of partial N\'eel states and Bethe states,
\href{http://dx.doi.org/10.1088/1742-5468/2016/02/023107}{J. Stat. Mech. (2016) 23107}.

\bibitem{dkm-16}
M. de Leeuw, C. Kristjansen, and S. Mori, AdS/dCFT one-point functions of the SU(3) sector,
\href{http://dx.doi.org/10.1016/j.physletb.2016.10.044}{Phys. Lett. B {\bf 763}, 197 (2016)}.

\bibitem{hst-17}
D. X. Horv\'ath, S. Sotiriadis, and G. Tak\'acs, Initial states in integrable quantum field theory quenches from an integral equation hierarchy,
\href{http://dx.doi.org/10.1016/j.nuclphysb.2015.11.025}{Nucl. Phys. B {\bf 902}, 508 (2016)};\\
D. X. Horv\'ath and G. Tak\'acs, Overlaps after quantum quenches in the sine-Gordon model,
\href{	http://dx.doi.org/10.1016/j.physletb.2017.05.087}{Phys. Lett. B {\bf 771}, 539 (2017)}.

\bibitem{msca-16}
P. P. Mazza, J.-M. St\'{e}phan, E. Canovi, V. Alba, M. Brockmann, and M. Haque, 
Overlap distributions for quantum quenches in the anisotropic Heisenberg chain,
\href{http://dx.doi.org/10.1088/1742-5468/2016/01/013104}{J. Stat. Mech. (2016) 013104}.


\bibitem{pvc-16} 
L. Piroli, E. Vernier, and P. Calabrese, Exact steady states for quantum quenches in integrable Heisenberg spin chains,
\href{http://dx.doi.org/10.1103/PhysRevB.94.054313}{Phys. Rev. B {\bf 94}, 054313 (2016)}.

\bibitem{garrison-2015}
J.~R.~Garrison and T.~Grover, Does a single eigenstate encode the full Hamiltonian?, 
\href{http://arxiv.org/abs/1503.00729}{arXiv:1503.00729}. 

\bibitem{delfino-14}
G. Delfino, Quantum quenches with integrable pre-quench dynamics,
\href{https://doi.org/10.1088/1751-8113/47/40/402001}{J. Phys. A {\bf 47} (2014) 402001}.

\bibitem{btc-17}
B. Bertini, E. Tartaglia, and P. Calabrese, 
Quantum Quench in the Infinitely Repulsive Hubbard Model: The Stationary State, 
\href{http://arxiv.org/abs/1707.01073}{arXiv:1707.01073}.

\bibitem{mc-12}
J. Mossel and J.-S. Caux, Generalized TBA and generalized Gibbs,
\href{https://doi.org/10.1088/1751-8113/45/25/255001}{J. Phys. A {\bf  45}, 255001 (2012)}.





\end{thebibliography}
\end{document}